\message
{JNL.TEX version 0.95 as of 5/13/90.  Using CM fonts.}

\catcode`@=11
\expandafter\ifx\csname inp@t\endcsname\relax\let\inp@t=\input
\def\input#1 {\expandafter\ifx\csname #1IsLoaded\endcsname\relax
\inp@t#1%
\expandafter\def\csname #1IsLoaded\endcsname{(#1 was previously loaded)}
\else\message{\csname #1IsLoaded\endcsname}\fi}\fi
\catcode`@=12

\font\twelverm=cmr12			\font\twelvei=cmmi12
\font\twelvesy=cmsy10 scaled 1200	\font\twelveex=cmex10 scaled 1200
\font\twelvebf=cmbx12			\font\twelvesl=cmsl12
\font\twelvett=cmtt12			\font\twelveit=cmti12
\font\twelvesc=cmcsc10 scaled 1200	\font\twelvesf=cmss12
\font\twelvemib=cmr10 scaled 1200
                     
\font\tenmib=cmr10
\font\eightmib=cmr10 scaled 800 

\skewchar\twelvei='177			\skewchar\twelvesy='60
\skewchar\twelvemib='177

\newfam\mibfam

\def\twelvepoint{\normalbaselineskip=12.4pt plus 0.1pt minus 0.1pt
  \abovedisplayskip 12.4pt plus 3pt minus 9pt
  \belowdisplayskip 12.4pt plus 3pt minus 9pt
  \abovedisplayshortskip 0pt plus 3pt
  \belowdisplayshortskip 7.2pt plus 3pt minus 4pt
  \smallskipamount=3.6pt plus1.2pt minus1.2pt
  \medskipamount=7.2pt plus2.4pt minus2.4pt
  \bigskipamount=14.4pt plus4.8pt minus4.8pt
  \def\rm{\fam0\twelverm}          \def\it{\fam\itfam\twelveit}%
  \def\sl{\fam\slfam\twelvesl}     \def\bf{\fam\bffam\twelvebf}%
  \def\mit{\fam 1}                 \def\cal{\fam 2}%
  \def\sc{\twelvesc}		   \def\tt{\twelvett}%
  \def\sf{\twelvesf}               \def\mib{\fam\mibfam\twelvemib}%
  \textfont0=\twelverm   \scriptfont0=\tenrm   \scriptscriptfont0=\sevenrm
  \textfont1=\twelvei    \scriptfont1=\teni    \scriptscriptfont1=\seveni
  \textfont2=\twelvesy   \scriptfont2=\tensy   \scriptscriptfont2=\sevensy
  \textfont3=\twelveex   \scriptfont3=\twelveex\scriptscriptfont3=\twelveex
  \textfont\itfam=\twelveit
  \textfont\slfam=\twelvesl
  \textfont\bffam=\twelvebf \scriptfont\bffam=\tenbf
                            \scriptscriptfont\bffam=\sevenbf
  \textfont\mibfam=\twelvemib \scriptfont\mibfam=\tenmib
                              \scriptscriptfont\mibfam=\eightmib
  \normalbaselines\rm}


\mathchardef\alpha="710B
\mathchardef\beta="710C
\mathchardef\gamma="710D
\mathchardef\delta="710E
\mathchardef\epsilon="710F
\mathchardef\zeta="7110
\mathchardef\eta="7111
\mathchardef\theta="7112
\mathchardef\iota="7113
\mathchardef\kappa="7114
\mathchardef\lambda="7115
\mathchardef\mu="7116
\mathchardef\nu="7117
\mathchardef\xi="7118
\mathchardef\pi="7119
\mathchardef\rho="711A
\mathchardef\sigma="711B
\mathchardef\tau="711C
\mathchardef\phi="711E
\mathchardef\chi="711F
\mathchardef\psi="7120
\mathchardef\omega="7121
\mathchardef\varepsilon="7122
\mathchardef\vartheta="7123
\mathchardef\varpi="7124
\mathchardef\varrho="7125
\mathchardef\varsigma="7126
\mathchardef\varphi="7127


\def\beginlinemode{\endmode
  \begingroup\parskip=0pt \obeylines\def\\{\par}\def\endmode{\par\endgroup}}
\def\beginparmode{\endmode
  \begingroup \def\endmode{\par\endgroup}}
\let\endmode=\par
{\obeylines\gdef\
{}}
\def\singlespace{\baselineskip=\normalbaselineskip}

\def\oneandahalfspace{\baselineskip=\normalbaselineskip
  \multiply\baselineskip by 3 \divide\baselineskip by 2}
\def\doublespace{\baselineskip=\normalbaselineskip \multiply\baselineskip by 2}

\newcount\firstpageno
\firstpageno=2
\footline={\ifnum\pageno<\firstpageno{\hfil}\else{\hfil\twelverm\folio\hfil}\fi}
\def\toppageno{\global\footline={\hfil}\global\headline
  ={\ifnum\pageno<\firstpageno{\hfil}\else{\hfil\twelverm\folio\hfil}\fi}}
\let\rawfootnote=\footnote		
\def\footnote#1#2{{\rm\singlespace\parindent=0pt\parskip=0pt
  \rawfootnote{#1}{#2\hfill\vrule height 0pt depth 6pt width 0pt}}}
\def\raggedcenter{\leftskip=4em plus 12em \rightskip=\leftskip
  \parindent=0pt \parfillskip=0pt \spaceskip=.3333em \xspaceskip=.5em
  \pretolerance=9999 \tolerance=9999
  \hyphenpenalty=9999 \exhyphenpenalty=9999 }
\def\dateline{\rightline{\ifcase\month\or
  January\or February\or March\or April\or May\or June\or
  July\or August\or September\or October\or November\or December\fi
  \space\number\year}}
\def\received{\vskip 3pt plus 0.2fill
 \centerline{\sl (Received\space\ifcase\month\or
  January\or February\or March\or April\or May\or June\or
  July\or August\or September\or October\or November\or December\fi
  \qquad, \number\year)}}


\hsize=6.5truein
\hoffset=0pt
\vsize=8.9truein
\voffset=0pt
\parskip=\medskipamount
\def\\{\cr}
\twelvepoint		
\doublespace		
\overfullrule=0pt	




\def\title			
  {\null\vskip 3pt plus 0.2fill
   \beginlinemode \doublespace \raggedcenter \bf}

\def\author			
  {\vskip 3pt plus 0.2fill \beginlinemode
   \singlespace \raggedcenter\sc}

\def\affil			
  {\vskip 3pt plus 0.1fill \beginlinemode
   \oneandahalfspace \raggedcenter \sl}

\def\abstract			
  {\vskip 3pt plus 0.3fill \beginparmode
   \oneandahalfspace ABSTRACT: }

\def\endtitlepage		
  {\endpage			
   \body}
\let\endtopmatter=\endtitlepage

\def\body			
  {\beginparmode}		

\def\head#1{			
  \goodbreak\vskip 0.5truein	
  {\immediate\write16{#1}
   \raggedcenter \uppercase{#1}\par}
   \nobreak\vskip 0.25truein\nobreak}

\def\subhead#1{			
  \vskip 0.25truein		
  {\raggedcenter {#1} \par}
   \nobreak\vskip 0.25truein\nobreak}

\def\beginitems{
\par\medskip\bgroup\def\i##1 {\item{##1}}\def\ii##1 {\itemitem{##1}}
\leftskip=36pt\parskip=0pt}
\def\enditems{\par\egroup}

\def\beneathrel#1\under#2{\mathrel{\mathop{#2}\limits_{#1}}}

\def\refto#1{$^{#1}$}		

\def\references			
  {\head{References}		
   \beginparmode
   \frenchspacing \parindent=0pt \leftskip=1truecm
   \parskip=8pt plus 3pt \everypar{\hangindent=\parindent}}

\gdef\refis#1{\item{#1.\ }}			

\gdef\journal#1, #2, #3, 1#4#5#6{		
    {\sl #1~}{\bf #2}, #3 (1#4#5#6)}		

\def\endreferences{\body}

\def\figurecaptions		
  {\endpage
   \beginparmode
   \head{Figure Captions}
}

\def\endpage			
  {\vfill\eject}

\def\endpaper			
  {\endmode\vfill\supereject}


\def\heading				
  {\vskip 0.5truein plus 0.1truein	
   \beginparmode \def\\{\par} \parskip=0pt \singlespace \raggedcenter}

\def\subheading				
  {\vskip 0.25truein plus 0.1truein	
   \beginlinemode \singlespace \parskip=0pt \def\\{\par}\raggedcenter}

\def\tag#1$${\eqno(#1)$$}

\def\align#1$${\eqalign{#1}$$}

\def\aligntag#1$${\gdef\tag##1\\{&(##1)\cr}\eqalignno{#1\\}$$
  \gdef\tag##1$${\eqno(##1)$$}}

\def\endaligntag{}

\def\overset #1\to#2{{\mathop{#2}\limits^{#1}}}
\def\underset#1\to#2{{\let\next=#1\mathpalette\undersetpalette#2}}
\def\undersetpalette#1#2{\vtop{\baselineskip0pt
\ialign{$\mathsurround=0pt #1\hfil##\hfil$\crcr#2\crcr\next\crcr}}}


\def\ref#1{Ref.~#1}			
\def\Ref#1{Ref.~#1}			
\def\[#1]{[\cite{#1}]}
\def\cite#1{{#1}}
\let\eq=\Eq\let\eqs=\Eqs		
\def\(#1){(\call{#1})}
\def\call#1{{#1}}
\def\taghead#1{}
\def\frac#1#2{{#1 \over #2}}

\def\12{{1\over2}}

\def\sla{\raise.15ex\hbox{$/$}\kern-.57em}
\def\leaderfill{\leaders\hbox to 1em{\hss.\hss}\hfill}
\def\twiddle{\lower.9ex\rlap{$\kern-.1em\scriptstyle\sim$}}
\def\bigtwiddle{\lower1.ex\rlap{$\sim$}}
\def\gtwid{\mathrel{\raise.3ex\hbox{$>$\kern-.75em\lower1ex\hbox{$\sim$}}}}
\def\ltwid{\mathrel{\raise.3ex\hbox{$<$\kern-.75em\lower1ex\hbox{$\sim$}}}}
\def\square{\kern1pt\vbox{\hrule height 1.2pt\hbox{\vrule width 1.2pt\hskip 3pt
   \vbox{\vskip 6pt}\hskip 3pt\vrule width 0.6pt}\hrule height 0.6pt}\kern1pt}
\def\tdot#1{\mathord{\mathop{#1}\limits^{\kern2pt\ldots}}}

\def\pmb#1{\setbox0=\hbox{#1}%
  \kern-.025em\copy0\kern-\wd0
  \kern  .05em\copy0\kern-\wd0
  \kern-.025em\raise.0433em\box0 }

\catcode`@=11
\newcount\r@fcount \r@fcount=0
\newcount\r@fcurr
\immediate\newwrite\reffile
\newif\ifr@ffile\r@ffilefalse
\def\w@rnwrite#1{\ifr@ffile\immediate\write\reffile{#1}\fi\message{#1}}

\def\writer@f#1>>{}
\def\referencefile{
  \r@ffiletrue\immediate\openout\reffile=\jobname.ref%
  \def\writer@f##1>>{\ifr@ffile\immediate\write\reffile%
    {\noexpand\refis{##1} = \csname r@fnum##1\endcsname = %
     \expandafter\expandafter\expandafter\strip@t\expandafter%
     \meaning\csname r@ftext\csname r@fnum##1\endcsname\endcsname}\fi}%
  \def\strip@t##1>>{}}

\def\citeall#1{\xdef#1##1{#1{\noexpand\cite{##1}}}}
\def\cite#1{\each@rg\citer@nge{#1}}	

\def\each@rg#1#2{{\let\thecsname=#1\expandafter\first@rg#2,\end,}}
\def\first@rg#1,{\thecsname{#1}\apply@rg}	
\def\apply@rg#1,{\ifx\end#1\let\next=\relax
\else,\thecsname{#1}\let\next=\apply@rg\fi\next}

\def\citer@nge#1{\citedor@nge#1-\end-}	
\def\citer@ngeat#1\end-{#1}
\def\citedor@nge#1-#2-{\ifx\end#2\r@featspace#1 
  \else\citel@@p{#1}{#2}\citer@ngeat\fi}	
\def\citel@@p#1#2{\ifnum#1>#2{\errmessage{Reference range #1-#2\space is bad.}%
    \errhelp{If you cite a series of references by the notation M-N, then M and
    N must be integers, and N must be greater than or equal to M.}}\else%
 {\count0=#1\count1=#2\advance\count1 by1\relax\expandafter\r@fcite\the\count0,%
  \loop\advance\count0 by1\relax
    \ifnum\count0<\count1,\expandafter\r@fcite\the\count0,%
  \repeat}\fi}

\def\r@featspace#1#2 {\r@fcite#1#2,}	
\def\r@fcite#1,{\ifuncit@d{#1}
    \newr@f{#1}%
    \expandafter\gdef\csname r@ftext\number\r@fcount\endcsname%
                     {\message{Reference #1 to be supplied.}%
                      \writer@f#1>>#1 to be supplied.\par}%
 \fi%
 \csname r@fnum#1\endcsname}
\def\ifuncit@d#1{\expandafter\ifx\csname r@fnum#1\endcsname\relax}%
\def\newr@f#1{\global\advance\r@fcount by1%
    \expandafter\xdef\csname r@fnum#1\endcsname{\number\r@fcount}}

\let\r@fis=\refis			
\def\refis#1#2#3\par{\ifuncit@d{#1}
   \newr@f{#1}%
   \w@rnwrite{Reference #1=\number\r@fcount\space is not cited up to now.}\fi%
  \expandafter\gdef\csname r@ftext\csname r@fnum#1\endcsname\endcsname%
  {\writer@f#1>>#2#3\par}}

\def\ignoreuncited{
   \def\refis##1##2##3\par{\ifuncit@d{##1}%
     \else\expandafter\gdef\csname r@ftext\csname r@fnum##1\endcsname\endcsname%
     {\writer@f##1>>##2##3\par}\fi}}

\def\r@ferr{\endreferences\errmessage{I was expecting to see
\noexpand\endreferences before now;  I have inserted it here.}}
\let\r@ferences=\references
\def\references{\r@ferences\def\endmode{\r@ferr\par\endgroup}}

\let\endr@ferences=\endreferences
\def\endreferences{\r@fcurr=0
  {\loop\ifnum\r@fcurr<\r@fcount
    \advance\r@fcurr by 1\relax\expandafter\r@fis\expandafter{\number\r@fcurr}%
    \csname r@ftext\number\r@fcurr\endcsname%
  \repeat}\gdef\r@ferr{}\endr@ferences}


\let\r@fend=\endpaper\gdef\endpaper{\ifr@ffile
\immediate\write16{Cross References written on []\jobname.REF.}\fi\r@fend}

\catcode`@=12

\citeall\refto		
\citeall\ref		%
\citeall\Ref		%

\catcode`@=11
\newcount\tagnumber\tagnumber=0

\immediate\newwrite\eqnfile
\newif\if@qnfile\@qnfilefalse
\def\write@qn#1{}
\def\writenew@qn#1{}
\def\w@rnwrite#1{\write@qn{#1}\message{#1}}
\def\@rrwrite#1{\write@qn{#1}\errmessage{#1}}

\def\taghead#1{\gdef\t@ghead{#1}\global\tagnumber=0}
\def\t@ghead{}

\expandafter\def\csname @qnnum-3\endcsname
  {{\t@ghead\advance\tagnumber by -3\relax\number\tagnumber}}
\expandafter\def\csname @qnnum-2\endcsname
  {{\t@ghead\advance\tagnumber by -2\relax\number\tagnumber}}
\expandafter\def\csname @qnnum-1\endcsname
  {{\t@ghead\advance\tagnumber by -1\relax\number\tagnumber}}
\expandafter\def\csname @qnnum0\endcsname
  {\t@ghead\number\tagnumber}
\expandafter\def\csname @qnnum+1\endcsname
  {{\t@ghead\advance\tagnumber by 1\relax\number\tagnumber}}
\expandafter\def\csname @qnnum+2\endcsname
  {{\t@ghead\advance\tagnumber by 2\relax\number\tagnumber}}
\expandafter\def\csname @qnnum+3\endcsname
  {{\t@ghead\advance\tagnumber by 3\relax\number\tagnumber}}

\def\equationfile{%
  \@qnfiletrue\immediate\openout\eqnfile=\jobname.eqn%
  \def\write@qn##1{\if@qnfile\immediate\write\eqnfile{##1}\fi}
  \def\writenew@qn##1{\if@qnfile\immediate\write\eqnfile
    {\noexpand\tag{##1} = (\t@ghead\number\tagnumber)}\fi}
}

\def\callall#1{\xdef#1##1{#1{\noexpand\call{##1}}}}
\def\call#1{\each@rg\callr@nge{#1}}

\def\each@rg#1#2{{\let\thecsname=#1\expandafter\first@rg#2,\end,}}
\def\first@rg#1,{\thecsname{#1}\apply@rg}
\def\apply@rg#1,{\ifx\end#1\let\next=\relax%
\else,\thecsname{#1}\let\next=\apply@rg\fi\next}

\def\callr@nge#1{\calldor@nge#1-\end-}
\def\callr@ngeat#1\end-{#1}
\def\calldor@nge#1-#2-{\ifx\end#2\@qneatspace#1 %
  \else\calll@@p{#1}{#2}\callr@ngeat\fi}
\def\calll@@p#1#2{\ifnum#1>#2{\@rrwrite{Equation range #1-#2\space is bad.}
\errhelp{If you call a series of equations by the notation M-N, then M and
N must be integers, and N must be greater than or equal to M.}}\else%
 {\count0=#1\count1=#2\advance\count1 by1\relax\expandafter\@qncall\the\count0,%
  \loop\advance\count0 by1\relax%
    \ifnum\count0<\count1,\expandafter\@qncall\the\count0,%
  \repeat}\fi}

\def\@qneatspace#1#2 {\@qncall#1#2,}
\def\@qncall#1,{\ifunc@lled{#1}{\def\next{#1}\ifx\next\empty\else
  \w@rnwrite{Equation number \noexpand\(>>#1<<) has not been defined yet.}
  >>#1<<\fi}\else\csname @qnnum#1\endcsname\fi}

\let\eqnono=\eqno
\def\eqno(#1){\tag#1}
\def\tag#1$${\eqnono(\displayt@g#1 )$$}

\def\aligntag#1\endaligntag
  $${\gdef\tag##1\\{&(##1 )\cr}\eqalignno{#1\\}$$
  \gdef\tag##1$${\eqnono(\displayt@g##1 )$$}}

\def\eqalignno#1{\displ@y \tabskip\centering
  \halign to\displaywidth{\hfil$\displaystyle{##}$\tabskip\z@skip
    &$\displaystyle{{}##}$\hfil\tabskip\centering
    &\llap{$\displayt@gpar##$}\tabskip\z@skip\crcr
    #1\crcr}}

\def\displayt@gpar(#1){(\displayt@g#1 )}

\def\displayt@g#1 {\rm\ifunc@lled{#1}\global\advance\tagnumber by1
        {\def\next{#1}\ifx\next\empty\else\expandafter
        \xdef\csname @qnnum#1\endcsname{\t@ghead\number\tagnumber}\fi}%
  \writenew@qn{#1}\t@ghead\number\tagnumber\else
        {\edef\next{\t@ghead\number\tagnumber}%
        \expandafter\ifx\csname @qnnum#1\endcsname\next\else
        \w@rnwrite{Equation \noexpand\tag{#1} is a duplicate number.}\fi}%
  \csname @qnnum#1\endcsname\fi}

\def\ifunc@lled#1{\expandafter\ifx\csname @qnnum#1\endcsname\relax}

\let\@qnend=\end\gdef\end{\if@qnfile
\immediate\write16{Equation numbers written on []\jobname.EQN.}\fi\@qnend}

\catcode`@=12


\def\CM{{\cal{M}}}
\def\str{{\rm str}}
\def\Mhat{{\hat M}}

{\parindent=0pt November 1997 \hfill{Wash. U. HEP/97-68 }}
\rightline{hep-lat/9711033}

\title Applications of Partially Quenched Chiral Perturbation Theory

\author Maarten F.L. Golterman${}^{1}$%
\footnote{}{${}^{1}$ e-mail: maarten@aapje.wustl.edu}%
\ and Ka Chun Leung${}^{2}$%
\footnote{}{${}^{2}$ e-mail: leung@hbar.wustl.edu}%
\affil Department of Physics 
       Washington University 
       St. Louis, MO 63130, USA

\abstract{Partially quenched theories are theories in which the
valence- and sea-quark masses are different.
In this paper we calculate the nonanalytic 
one-loop corrections of some physical 
quantities: the chiral condensate, weak decay 
constants, Goldstone boson masses, 
$B_K$ and the $K^+ \rightarrow \pi^+ \pi^0$ decay 
amplitude, using partially quenched chiral perturbation theory. 
Our results for weak decay constants and masses agree with, and
generalize, results of previous work by Sharpe. 
We compare $B_K$ and the $K^+$ decay amplitude with their real-world 
values in some examples. 
For the latter quantity,  two other systematic
effects that plague lattice computations, namely, finite-volume effects and
unphysical values of the
quark masses and pion external momenta 
are also considered. We find that typical one-loop corrections can be 
substantial.}

\endtopmatter

\subhead{\bf 1. Introduction}

Recently, large scale numerical lattice QCD computations have started to
move away from the quenched approximation by taking virtual-quark loop 
effects into account. However, since the computation of quark determinants
is much more expensive than that of quark propagators, one is often
restricted to only very few values of the sea-quark mass, while,
given an ensemble of ``dynamical fermion" gauge configurations, one
can explore a much larger range of values for the valence-quark masses. 
This naturally leads us to consider partially quenched theories, which
are theories in which the valence- and sea-quark masses are not equal.

Chiral Perturbation Theory (ChPT) plays an important role in the analysis
of results from lattice computations (see for instance ref.
[\cite{Sharpe96}]).  Since the quenched or partially
quenched approximations of lattice QCD are theories different from the
full unquenched theory, ChPT needs to be adapted to these situations.
For the quenched case, this has been extensively investigated (for 
a rather complete list of references, see ref. [\cite{MGMainz}]),
but much less work has been done for the partially quenched case.

Partially Quenched Chiral Perturbation Theory (PQChPT) was developed in ref.
[\cite{BG}], and used to calculate one-loop expressions for Goldstone boson
masses and decay constants in ref. [\cite{SharpePQ}].  In this paper, we
revisit Goldstone boson masses and decay constants, and we also give
the one-loop expressions for the chiral condensate, $B_K$ and the
$K^+\rightarrow\pi^+\pi^0$ decay amplitude. (For an application to
heavy-light decay constants and B-parameters, see ref. [\cite{SZ}].)

In the partially quenched case (unlike in the quenched case), the $\eta'$
meson is heavy in the sense that its mass does not vanish in the chiral
limit.  One can therefore approach PQChPT in the same way as one does in
the unquenched case, where the $\eta'$ is integrated out, and only the
Goldstone bosons are kept in the effective lagrangian.  This is the
approach taken in ref. [\cite{SharpePQ}].  In the completely quenched
case, however, the $\eta'$ is light, and needs to be kept in the
effective lagrangian, along with the other Goldstone bosons. The theory
depends on a mass scale $m_0$, independent of the quark masses, 
which in the unquenched theory is the
singlet part of the $\eta'$ mass, but in the quenched theory no longer
appears in any pole mass [\cite{KS,qflow,BGQ,SharpeBK}]. A systematic 
expansion in the quenched case only exists if we treat the ratio of
$m_0$ to the chiral-symmetry breaking scale as an independent {\it small}
parameter, in addition to the light quark masses, and if we stay away
from the chiral limit, where infrared divergences occur 
[\cite{BGQ,SharpeBK,BGscat,CP}].

In partially quenched lattice computations, it may 
happen that the scale set by the sea-quark mass is not small compared to
the scale $m_0$. In this paper we therefore keep the $\eta'$ in the
effective theory, and show that one can still develop PQChPT systematically
if one assumes, as in QChPT, that the ratio of $m_0$ to the
chiral-symmetry breaking scale is reasonably small. This generalizes
the results of ref. [\cite{SharpePQ}]. 

The outline of this paper is as follows.  In section 2, we give a quick
review of PQChPT, and discuss the role of the $\eta'$ in more detail.
We explain more precisely how our calculation generalizes that of
ref. [\cite{SharpePQ}].  We then calculate in section 3 the chiral condensate 
and Goldstone boson masses and decay constants to one loop as a function of
the quark masses, and investigate their dependence on $m_0$. 
We restrict ourselves to the theory with $N$ degenerate sea quarks. 
In section 4, we calculate $B_K$ for non-degenerate valence-quark masses
and the $K^+\rightarrow\pi^+\pi^0$ decay amplitude for degenerate 
valence-quark masses. For $K^+\rightarrow\pi^+\pi^0$ we also discuss 
other systematic errors which have affected lattice computations of this
quantity to date.  In section 5, we give numerical examples of the role 
of one-loop corrections for $B_K$ and $K^+\rightarrow\pi^+\pi^0$ in a 
comparison between the real world and the partially quenched theory with
parameters typical of a lattice computation.  
More details on the assumptions underlying these numerical examples
will be given in due course. We end with our conclusions. 

\subhead{\bf 2. Essentials of Partially Quenched Chiral Perturbation Theory}

Consider a QCD-like theory with $n$ flavors of quarks $q_i$, 
each with arbitrary mass $m_i$. We then partially quench the theory by adding 
$n-N$ flavors of (unphysical) quarks $\tilde{q}_i$ 
obeying bosonic statistics [\cite{Morel}], which we will call ghost-quarks. 
The ghost-quark masses are 
chosen to equal those of the first $n-N$ physical quarks. These $n-N$ 
quarks  are then quenched since their loop contributions are exactly canceled 
by their bosonic 
counterparts. The remaining $N$ quarks contribute through (sea-)quark loops 
to physical quantities. With the $n-N$ quarks identified as valence quarks, the 
theory 
corresponds to partially quenched QCD with $N$ 
flavors of sea quarks. Recent partially quenched lattice computations 
use degenerate sea-quark masses, hence, in the following, 
we will set all sea-quark masses $m_i$, $i=n-N+1,\cdots,n$, equal to $m_S$.

The full chiral symmetry of the theory is the semi-direct product of 
graded groups  
$G\equiv [SU(n|n-N)\otimes SU(n|n-N)]\bigcirc\kern -0.43cm s\; U(1)$ 
after the  anomaly has been taken in account [\cite{BGQ}]. 
We will briefly discuss the construction of the lagrangian for PQChPT
for our choice of quark masses.

The unitary field $\Sigma$ is defined through
$$\Sigma \equiv \exp{(2i\Phi/f)} \eqno(Sigma)$$
from the $(2n-N)\times(2n-N)$ hermitian matrix field 
$$\Phi=\pmatrix{\phi&\chi^\dagger\cr\chi&{\tilde{\phi}}\cr}\ ,\eqno(Phi)$$
where $\phi$ is the $n\times n$ matrix of ordinary mesons made from the $n$ 
ordinary quarks and antiquarks, $\tilde \phi$ is the corresponding 
$(n-N)\times(n-N)$ matrix for ghost-quark mesons, and $\chi$ is an $(n-N)\times 
n$ matrix of fermionic
mesons made from a ghost-quark and an ordinary antiquark. $f$ is the 
tree-level pion decay constant. The $(2n-N)\times(2n-N)$ quark-mass matrix 
$\Mhat$ is defined by $\Mhat_{ij}=m_i \delta_{ij}$, 
where, as already discussed, 
the masses $m_i$ for $i=1,\cdots , n-N$ are arbitrary, 
and equal to the ghost-quark masses, $i=n+1, \cdots ,2n-N$. 
The remaining $N$ masses $m_i$, $i=n-N+1,\cdots,n$, are degenerate and equal to 
$m_S$.

As discussed in refs. [\cite{BGQ,BG}], 
the super-$\eta'$ field $\Phi_0\equiv \str(\Phi)$ 
is invariant under the full chiral group $G$ and introduces 
new parameters $m^2_0$ and $\alpha$ 
into the $O(p^2)$ euclidean lagrangian for PQChPT, which reads
$${\cal L}={f^2\over 8}\; \str\left( \partial_\mu \Sigma \partial_\mu 
\Sigma^\dagger \right)
-{f^2\mu\over 4}\;\str\left(\Mhat\Sigma+\Mhat\Sigma^\dagger\right)+{m^2_0\over 
6}\; \Phi^2_0 +{\alpha\over 6} 
\left( \partial_\mu \Phi_0 \right)\left(\partial_\mu \Phi_0 \right).
\eqno(Lag)$$
We define 
$$M^2_{ij} \equiv\mu (m_i +m_j ) \eqno(M2ij)$$
(for $i=j=n-N+1,\cdots,n$, 
this simplifies to $M^2_{ii}=2\mu m_S \equiv M^2_{SS}$). 
It is instructive to display the two-point functions for the neutral
mesons ({\it i.e.} the diagonal fields $\Phi_{ii}$) explicitly. 
In the diagonal 
basis with states $\Phi_{ii}$, $i=1,\cdots, 2n-N$, corresponding to 
$q_1 {\overline q}_1 ,\  q_2 {\overline q}_2 , \cdots,$ and their 
ghost-quark counterparts, these two-point functions are, in momentum 
space [\cite{BG}], 
$$G_{i,j}(p)={\delta_{ij}\epsilon_i\over{p^2+M^2_{ii}}}-
{1\over{(3+N\alpha)}}{{(m^2_0+\alpha p^2) 
(p^2 +M^2_{SS} )} \over 
{(p^2 +M^2_{ii}) (p^2+M^2_{jj})(p^2 +m^2_{\eta'})} }\ ,\eqno(Gij)$$
where 
$$\epsilon_i =\cases{ +1,&for $1 \le i \le n$\cr
                                    -1,&for $n+1 \le i \le 2n-N$\cr}\ , 
\eqno(ep)$$
and
$$m^2_{\eta'} ={M^2_{SS} +N m^2_0 /3\over 1+N\alpha /3}\ ,\eqno(MPhi)$$
which is the square of the pole mass for the $\Phi_0$ two-point function,
as can easily be verified from \eq{Gij}. It is equal to the square of  
the ``$\eta'$" meson mass in the 
$SU(N)$ theory constructed only from the unquenched quarks [\cite{BG}]. 
It does not vanish in the chiral limit $m_{S} \rightarrow 0$. 

A simplification occurs in partially quenched QCD. 
For a correlation function in which only $k$ 
out of the $n-N$ valence quarks are on the  
external lines, the rest of the $n-N-k$ valence 
quarks do not contribute at all, and therefore this correlation function
does not depend on their masses. In most of this paper, we will
only consider quantities involving at most two valence quarks, 
{\it i.e.} we will only consider valence quarks $q_i$ with
$i=1,2$. 

Compared to quenched ChPT (QChPT), PQChPT introduces an extra mass scale
$M_{SS}$.  We can therefore consider various possible ``expansion schemes." 
Let $\Lambda_{p}$ be the cutoff of the partially quenched theory. The 
first scheme, adopted in ref. [\cite{SharpePQ}], 
takes $M_{11},\ M_{22},\ M_{SS}<m_{\eta'}\sim\Lambda_{p}$. 
This corresponds to the usual set-up for power counting in ChPT, in which
the $\eta'$ is not a Goldstone boson but instead a ``heavy" particle,
due to the presence of the scale
$m_0$, which does not vanish in the chiral limit.  In this scheme, 
the field $\Phi_0$ can be integrated out, absorbing dependence on the 
parameters $\alpha$ and $m_0^2$ into 
$O(p^4)$ terms in the chiral lagrangian.
This can be illustrated by considering for instance 
the two-point function $G_{1,1}$ ({\it cf.} \eq{Gij}). 
We first write the second term in $G_{1,1}$ as 
$$ -{m^2_0 +\alpha p^2 \over 3+N\alpha } \left( 
{m^2_{\eta'} -M^2_{SS} \over (m^2_{\eta'} -M^2_{11} )^2} 
\left[ {1\over p^2 +M^2_{11} }-{1\over {p^2 +m^2_{\eta'}}} \right]
+{M^2_{SS}-M^2_{11} \over m^2_{\eta'} -M^2_{11} }{1\over (p^2 +M^2_{11} )^2} 
\right)\ .
\eqno(G11)
$$
For large $m_{\eta'}$, the expansion of the term containing the pole 
$1/ (p^2 +m^2_{\eta'} )$ in powers of $M^2_{11} /m^2_{\eta'} $ and 
$M^2_{SS} / m^2_{\eta'}$ generates contact terms which can be absorbed 
into $O(p^4)$ coefficients of the chiral lagrangian. Similarly expanding 
the remaining terms in \eq{G11} gives the leading terms
$$ -{1\over N} \left( {1\over p^2 +M^2_{11} } +{M^2_{SS} -M^2_{11} \over (p^2 
+M^2_{11} )^2 } \right) \ . \eqno(GG11)$$ 
The poles in this expression, which no longer depends on $\alpha$ and
$m_0^2$, lead to nonanalytic terms at one loop.
For  details and further discussion, see ref. [\cite{SharpePQ}]. 

In QChPT, there is no heavy particle with a mass determined by the
parameter $m_0^2$, as can been seen by setting $N=0$ in 
\eq{Gij}. (The $\Phi_0$ two-point function vanishes for $N=0$
[\cite{BGQ}].) The second term in \eq{Gij} is proportional to $m_0^2+
\alpha p^2$, and has a double pole at the Goldstone meson masses
$M_{ii}^2$.  (Of course, no $M_{SS}$ dependence remains.)
Hence, this term needs to be kept in the quenched
approximation, and the decoupling as described above does not occur.
It turns out that, in order to make sense of the chiral expansion, we need
to assume that the parameter $(m_0^2/3)/(4\pi f_\pi)^2$ is small,
in addition to the usual requirement that $M_{ii}^2/(4\pi f_\pi)^2$
be small [\cite{BGQ,SharpeBK}]. QChPT therefore corresponds to a 
second ``scheme" in which 
$M_{11} ,\ M_{22} ,\ m_0 < 
\Lambda_{p}$ and $M_{SS} > \Lambda_{p}$ which corresponds to 
freezing out sea-quark loop effects, effectively setting $N=0$.  
We note that in the unquenched
theory $(m_0^2/3)/(4\pi f_\pi)^2\approx 0.09$ (where $f_\pi=132$~MeV is the
physical pion decay constant).  
It is believed that the value of $m_0$ is not very different
in the quenched theory [\cite{Sharpe96}].

Since the partially quenched theory
``interpolates" between the quenched and unquenched theories, it is
natural to consider a third scheme, which we will now explain.
First, as in ref. [\cite{SharpePQ}], we take $M_{11},\ M_{22},\ M_{SS}
<\Lambda_p$,
and $M_{11},\ M_{22}<m_{\eta'}$.  However, we would like to leave the 
ratio of $M_{SS}$ and $m_0$ arbitrary.  Let us consider the possibilities. 
If $M_{SS}<m_{\eta'}$ we can systematically study the effective theory
for mesons with masses $M_{11},\ M_{22}$ and $M_{SS}$ by integrating
out the $\eta'$, because now $M_{11},\ M_{22},\ M_{SS}<m_{\eta'}$, 
as in ref. [\cite{SharpePQ}].  If, on the contrary, $M_{SS}$ is of order
$m_0$ or larger, this would be incorrect.  Now, we can only systematically
investigate the effective theory for mesons with masses $M_{11},\ M_{22}$ and
$M_{SS}$, if we keep the $\eta'$ in the effective theory, and, as in the
completely quenched case, assume that $m_0<\Lambda_p$.

In partially quenched lattice computations, often $M_{SS}$ is fairly large,
and we may well be in the situation that indeed $M_{SS}$ is of order $m_0$.  
Therefore, in this paper, we will be interested in calculating the nonanalytic
dependence on the quark masses $M_{11},\ M_{22},\ M_{SS}$
for various quantities at one loop, taking
$M_{11},\ M_{22},\ M_{SS},\ m_0<\Lambda_p$ and $M_{11},\ M_{22}<m_{\eta'}$.
Note that we can still consider the case that $M_{SS}<m_0$ if 
$M_{11},\ M_{22}<m_0$, which leads us back to the assumptions 
made in ref. [\cite{SharpePQ}].  Indeed, when we expand our results 
for meson masses and decay constants in
$M_{SS}/m_{\eta'}$, we obtain those of ref. [\cite{SharpePQ}].%
\footnote *{Due to a difference in normalization, our tree-level
weak decay constant $f$ is by a factor $\sqrt{2}$
larger than that of refs. [\cite{SharpePQ,SharpeBK}]. This should be taken
into account before comparing results.}  On the other hand, if we take
$M_{SS}$ large and expand in $M_{11}/M_{SS}$, $M_{22}/M_{SS}$ and
$m_0/M_{SS}$, we obtain the quenched ({\it i.e.} $N=0$) results of
refs. [\cite{BGQ,SharpeBK,GL}].

Before we present our results, we should address one more issue.
In our scheme, there will be one-loop contributions proportional to
$\log{m^2_{\eta'}}$ coming from $\Phi_0$ tadpoles.  For values of the
meson masses which are small relative to $m_{\eta'}$,
such terms can be absorbed into the $O(p^4)$ coefficients (after
expanding in $M_{ii}/m_{\eta'}$).  However, this is not true for
$M_{SS}$ when $M_{SS}$ is of order $m_0$ or larger.  In this case,
the dependence of $\Phi_0$ tadpoles on $M_{SS}$ will be complicated.
Moreover, the contributions from $\Phi_0$ tadpoles will depend on other
$\eta'$ coupling constants (analogous to $m_0^2$ and $\alpha$)
which are basically unknown. (We did not include these couplings in
\eq{Lag}; for a more complete expression, see refs. 
[\cite{BGQ,BG}].)

Therefore, in this paper, we will take the point of view that we
keep $M_{SS}$ fixed, and calculate the nonanalytic dependence on
the valence-quark masses.  This, then, allows us to ignore 
contributions from $\Phi_0$ tadpoles, which we will do in the rest
of this paper.  It does not affect the coefficients of the other 
chiral logarithms. Note that we do not assume 
the ratios of the valence-quark masses and the sea-quark mass to be
small; our scheme allows for arbitrary values of these ratios.

\subhead{\bf 3. Chiral condensate, masses and weak 
decay constants}

We list the one-loop expressions for the condensate, meson masses and decay
constants for a theory with nondegenerate valence-quark masses 
$m_1$ and $m_2$ and sea-quark mass $m_S$
in terms of the bare parameters $f$, $M_{ij}^2$, $m_0^2$ and
$\alpha$, including only nonanalytic terms (with $M_{iS}^2=(M_{ii}^2+M_{SS}^2)/2$,
{\it cf.} \eq{M2ij}):
$$
\eqalignno{
\left[ m_i \langle {\overline q_i} q_i \rangle \right]_{\rm 1-loop}
&=-{M^2_{ii} f^2 \over 4}\Biggl( 1
-2N{M_{iS}^2 \over(4\pi f)^2}\log{M_{iS}^2 \over \Lambda_{p}^2}\cr
&\phantom{=-{M^2_{ii} f^2 \over 4}\Biggl( 1}
-{2\over 3(4\pi f)^2}\left[ \CM^2-AM^2_{ii}  +\left( \CM^2 -2AM^2_{ii}\right)
\log{M^2_{ii} \over \Lambda_{p}^2}\right]
\Biggr)\cr
&=-{1\over 4}[M^2_{ii}]_{\rm 1-loop} [f^2_{ii}]_{\rm 1-loop}\;,&(mqq)\cr
\left[ M^2_{12}\right]_{\rm 1-loop}&=
M^2_{12}
\Biggl( 1+{2\over 3(4\pi f)^2}\Biggl[{M^2_{11} \left( \CM^2 -AM^2_{11}\right) 
\over M^2_{22} -M^2_{11}} 
\log{M^2_{11}\over \Lambda^2_{p}} \cr
&\phantom{M^2_{12}
\Biggl( 1+{2\over 3(4\pi f)^2}\Biggl[}
-{M^2_{22} \left( \CM^2 -AM^2_{22} \right) \over M^2_{22} -
M^2_{11}} \log{M^2_{22}\over \Lambda^2_{p}}\Biggr]\Biggr)
\;,&(M2)\cr
 {\left[ f_{12} \right]_{\rm 1-loop} /f}&=
1-{N\over 2(4\pi f)^2}\left(M_{1S}^2\log{M_{1S}^2\over \Lambda_{p}^2} 
+M_{2S}^2\log{M_{2S}^2 \over \Lambda_{p}^2}\right)&(f12)\cr
&\phantom{1}+{1\over 3(4\pi f)^2} \left(
\CM^2 \left({1\over 2\epsilon}\log{1+\epsilon \over1-\epsilon} -1 \right)  
+AM^2_{12}
\left( 1-{1-\epsilon^2 \over 2\epsilon}\log{1+\epsilon \over 1-\epsilon}\right)
\right)\;,
}$$
where 
$$\epsilon \equiv {m_2 -m_1 \over m_2 +m_1}={M^2_{22}-M^2_{11}
\over M^2_{22} +M^2_{11}}\ .
\eqno(epsilon)$$
All dependence on $m_0^2$ and $\alpha$ is embodied in the quantities
$\CM^2$ and $A$, with
$$\eqalignno{
\CM^2 &\equiv\ \ \;{M^2_{SS} \ m^2_0 \over {M^2_{SS} +Nm^2_0 /3}}
\ \ \;=\;{{3y}\over{1+Ny}}\;M_{SS}^2\;,&(calM)\cr&\cr
A&\equiv\;{\alpha M^4_{SS}+Nm^4_0/3\over\left(M^2_{SS}+N m^2_0/3\right)^2}
\;=\;{{\alpha+3Ny^2}\over{(1+Ny)^2}}\;,
&(A)}$$
where we introduced the ratio $y\equiv(m_0^2/3)/M_{SS}^2$.
The results of ref. [\cite{SharpePQ}] (with $\alpha_4=\alpha_5=\alpha_6
=\alpha_8=0$ in Eqs.~(13--20) of ref. [\cite{SharpePQ}]) 
can easily be obtained by expanding
$\CM^2$ and $A$
in $M_{SS}^2/m_{\eta'}^2$ (or $M_{SS}^2/m_0^2$), and keeping the leading
order terms, $3M_{SS}^2/N$ and $3/N$, respectively. (Subleading terms are
of higher order in the chiral expansion, and can be dropped.) 
This corresponds mathematically to taking the limit $y\rightarrow\infty$. 
Let us comment briefly on this comparison between our results and those
of ref. [\cite{SharpePQ}].  In our case, we keep the $\eta'$, whereas
in ref. [\cite{SharpePQ}] the $\eta'$ is integrated out.  One would 
therefore in general expect that in order to ``match" the two theories,
we would need to adjust the bare parameters.  For the quantities
considered here, it turns out that, at one loop,
 all nontrivial adjustments come
from $\Phi_0$-tadpole contributions.  However, 
we did not have to, and hence did not, 
include such contributions, as was explained in 
section 2.  We conclude that, for the Goldstone meson masses and decay
constants, no adjustment is needed. 

The results of QChPT [\cite{BGQ,SharpeBK}] can be obtained
by taking $N=0$, for which $\CM^2=m^2_0$ and $A=\alpha$. 

With degenerate quark masses $m_V\equiv m_1 =m_2$ in \eq{M2} and 
\eq{f12}, we obtain 
$$M^2_{VV} \equiv \left[ M^2_{11} \right]_{\rm 1-loop} =
M^2_{11} \left( 1-{2\over 3(4\pi f)^2} \left[ \CM^2 -AM^2_{11} +\left( \CM^2 
-2AM^2_{11} 
\right) \log{M^2_{11} \over \Lambda^2_{p}}\right] \right)\;,
\eqno(M2VV)
$$ 
and 
$${f_{VV} \over f} \equiv {\left[ f_{11} \right]_{\rm 1-loop} \over f} 
=1-N
{M_{1S}^2 \over (4\pi f)^2}\log{M_{1S}^2 \over \Lambda^2_{p}}\;.
\eqno(fVV)
$$
As in QChPT, the ratio of $M^2_{VV}$ to its tree-level value,
$M^2_{VV}/M^2_{11}$, is singular, while $f_{VV}$ is regular in the chiral 
limit $m_V \rightarrow 0$. Moreover,  $f_{VV}$ does not depend on $m^2_0$ 
and $\alpha$. 

We can define $M^2_{VS}$ and $f_{VS}$ to one loop from  
\eqs{M2,f12} by replacing $m_{1}$ by $m_{V}$ and $m_{2}$ by $m_{S}$. 
In the chiral limit $m_V \rightarrow 0$, keeping $m_S$ fixed,
we have $\epsilon \rightarrow 
1$ and the nonanalytic functions multiplying $\CM^2$ and $A$ in the
expression for $f_{VS}$, 
{\it cf.} \eq{f12}, diverge. 
It is obvious from \eq{M2}, that the ratio of $M^2_{VS}$ to its
tree-level value is regular in the chiral limit, $m_{V} \rightarrow 0$.

Of course, all quantities considered here also receive analytic
contributions from $O(p^4)$ terms in the chiral lagrangian.  Since
they are obtained from tree-level diagrams with $O(p^4)$ vertices,
they do not explicitly depend on the parameters $m_0^2$ and $\alpha$, and
therefore are identical to those reported in ref. [\cite{SharpePQ}],
to which we refer for their explicit form.
In ref. [\cite{SharpePQ}] it was pointed out that the quantity
$$M^2_{VV} -M^2_{V'S}\ \ \ \ {\rm with\ \ \ \ } m_{V'} =2m_V -m_S\ ,
\eqno(M2D)$$
is independent of $O(p^4)$ coefficients.

Another quantity which is independent of $O(p^4)$ coefficients,
first introduced in ref. [\cite{BGF}] for the quenched case,  and also
considered in ref. [\cite{SharpePQ}] for the partially quenched case
at $y=\infty$, is
$$\left[{f_{12} \over \sqrt{f_{11} f_{22}}}\right]_{\rm 1-loop}\!\! =1
+{1\over 3(4\pi f)^2} \left(
\CM^2 \left({1\over 2\epsilon}\log{1+\epsilon \over1-\epsilon} -1 \right)
+AM^2_{12}
\left( 1-{1-\epsilon^2 \over 2\epsilon}\log{1+\epsilon \over
1-\epsilon}\right)
\right). \eqno(RBG)
$$
For any $m_S$, this quantity diverges when
$m_1 \rightarrow 0$ with $m_2$ fixed ({\it i.e.} for $\epsilon \rightarrow
1$).  Note that $[f_{12}]_{\rm 1-loop}$ 
is an even function of $\epsilon$ (because
of symmetry under the interchange $m_1\leftrightarrow m_2$), so that
$[\;{f_{12} / \sqrt{f_{11} f_{22}}}\;]_{\rm 1-loop}=1+O(\epsilon^2)$
for $\epsilon\rightarrow 0$. 

{}From \eqs{M2,f12,M2VV}, we see that the coefficients of the chiral
logarithms of the valence-quark mass depend on the ratio $y$ of the 
parameter $m_0^2$ and the sea-quark mass, through the quantities
${\cal M}^2$ and $A$ ({\it cf.} \eqs{calM,A}).  
{}From (partially) quenched 
lattice data, it is estimated that $m_0^2/3$ presumably has a value
$m_K^2/2\;{<\atop\sim}\;m_0^2/3\;{<\atop\sim}\;m_K^2$ ($m_K=496$~MeV
is the physical kaon mass) 
[\cite{Sharpe96,KV}]. Typical lattice computations have $N=2$ and 
$m_K^2\;{<\atop\sim}\;M_{SS}^2\;{<\atop\sim}\;2m_K^2$. These values of
$m_0^2$ and $M_{SS}^2$ correspond to $y$ ranging from $y\approx 1/4$ to
$y\approx 1$.
This leads to ${\cal M}^2/M_{SS}^2=1/2$ for $y=1/4$ and to 
${\cal M}^2/M_{SS}^2=1$ for $y=1$.  For $y\rightarrow\infty$ one obtains
${\cal M}^2/M_{SS}^2=3/2$. This shows that for relatively heavy sea quarks,
there is a clear dependence of the coefficient of the chiral logarithms
on $m_0^2$. 
(Experience with quenched lattice data [\cite{Sharpe96}] indicates that
it is hard to fit the chiral logarithms reliably,
partially because of the ``competition" of $O(p^4)$ coefficients.
This may make it difficult to see the $y$ dependence of
the chiral logarithms in practice.)

The quantity $A$ also has an effect on the coefficients of the chiral
logarithms, in particular for values of the valence-quark mass of order
of the sea-quark mass.
Taking again $N=2$, we find $A=(8\alpha+3)/18$ for $y=1/4$ and 
$A=(\alpha+6)/9$ for $y=1$, while $A=3/2$ for $y\rightarrow\infty$.  It is 
clear that $A$ is more sensitive to the value of $\alpha$ for smaller values
of $y$.   We note that for $m_V/m_S=1$, $M_{11}^2/m_{\eta'}^2=1/(1+Ny)$
(for $\alpha=0$), so that our results may not be reliable 
for smaller values of $y$.

\subhead{\bf 4. $B_K$ and $K^+ \rightarrow \pi^+ \pi^0$ decay}

In this section, we will generalize earlier quenched one-loop
calculations for $B_K$ [\cite{SharpeBK,SharpeTASI,GL}] and 
$K^+ \rightarrow \pi^+ \pi^0$ [\cite{GL}] to the partially quenched
case.
In the following, $u$, $d$ and $s$ denote valence quarks with masses 
$m_1$, $m_1$ and $m_2$ respectively. 
The kaon B-parameter $B_K$ is defined as (with $M_K$ the mass of the
$d\overline{s}$ meson, {\it i.e.} at tree level $M_K=M_{12}$)
$$B_K={{\langle {\overline K}^0 \vert ({\bar s}d{\bar s}d)_{LL} 
\vert K^0 \rangle} \over {8 \over 3}{f_K^2 M_K^2}}\ , \eqno(BK)$$
in which the four-quark operator is defined by
$$({\bar q_i} q_j {\bar q_k} q_l)_{LL} =({\bar q_{iL}} \gamma^\mu q_{jL})
({\bar q_{kL}} \gamma_\mu q_{lL})\ , \eqno(anycurcur)$$
where $q_L={1 \over 2}(1-\gamma_5)q$ is a left-handed quark field. The 
denominator in \eq{BK} is the matrix element 
${\langle {\overline K}^0 \vert ({\bar s}d{\bar s}d)_{LL} 
\vert K^0 \rangle}$ evaluated by vacuum saturation. 
The $O(p^2)$ weak-interaction operator in ChPT corresponding to  
$({\bar s}d{\bar s}d)_{LL}$, which  is the $\Delta S=2$ 
component of a 27-plet under $SU(3)_L$, is 
$$O'=\alpha_{\scriptscriptstyle 27} \; t^{ij}_{kl}({\Sigma \partial^\mu 
\Sigma^\dagger})_i^{\ k}({\Sigma \partial_\mu \Sigma^\dagger})_j^{\ l} \ , 
\eqno(OOprime)$$
where the tensor $t^{ij}_{kl} $ is defined by setting 
$$t^{22}_{33}=1\;,\eqno(Oprimets)$$
while all other components are equal to zero. 
The parameter $\alpha_{\scriptscriptstyle 27}$ 
is the only new $O(p^2)$-operator coefficient; 
its value is determined by QCD dynamics.
At tree level, one obtains [\cite{DGH}]
$$\langle {\overline K}^0 |O'|K^0 \rangle={8\alpha_{\scriptscriptstyle 27} 
\over f^2}M^2_K\; , \eqno(Oprimetree)$$
and therefore,
$$B_K = {3\alpha_{\scriptscriptstyle 27} \over f^4} \equiv B
\eqno(B)$$
($f_K =f$ at tree level).
Since the partially quenched theory is different from unquenched QCD,
the partially quenched value of the coefficient 
$\alpha_{\scriptscriptstyle 27}$ is in principle different from the QCD
value.  We will make this explicit by using a subscript 
or superscript $p$ to denote
bare parameters of the partially quenched theory, specifically,
$\alpha_{\scriptscriptstyle 27}^p$, $B^p$ and $f_p$.

The partially quenched
one-loop result for $B^{p}_K$ with non-degenerate quark masses, 
keeping only nonanalytic terms, is
$$\eqalignno{
&B^{p}_K=B^{p} \times &(QBK)\cr
&\Biggl(1+
{M^2_{12} \over (4\pi f_{p})^2} \left( -2(3+\epsilon^2)\log{M^2_{12}\over 
\Lambda^2_{p}} -
(2+\epsilon^2)\log{(1-\epsilon^2)}
-3\epsilon\log{1+\epsilon\over 1-\epsilon}\right)\cr
&\phantom{\Biggl(1}
+{2\over 3(4\pi f_{p})^2} \Biggl[
\CM^2 \left({2-\epsilon^2 \over 2\epsilon}\log{1+\epsilon\over 1-\epsilon} 
-2\right)\cr
&\phantom{\Biggl(1++{2\over 3(4\pi f_{p})^2} }
+AM^2_{12} \left(
2+\epsilon^2 -{1-2\epsilon^2-\epsilon^3\over \epsilon}\log{1+\epsilon\over 
1-\epsilon}
+2\epsilon^2 \log{\left( {M^2_{12} \left( 1-\epsilon \right) \over 
\Lambda^2_{p}} \right)}\right)
\Biggr]
\Biggr)\ , 
}$$
where $\epsilon$ is defined in \eq{epsilon}. It can easily be seen that \eq{QBK} 
does not depend
on $m^2_0$ and $\alpha$ in the case of degenerate quark masses 
($\epsilon =0$), 
just as in QChPT. Actually, apart from changing the values of $B$ and $f$ to 
their  (partially) quenched values, (partial) quenching 
does not introduce any 
change in the nonanalytic one-loop corrections of $B_K$ in the degenerate 
case.  For the quenched case this was already discussed in ref.
[\cite{SharpeBK}], and the fact that this is also true in PQChPT does not
come as a surprise, since PQChPT is ``in between" ChPT and QChPT (see also
the discussion of $K^+ \rightarrow \pi^+ \pi^0$ below).  In the
nondegenerate case, the partially quenched result follows from the quenched
result by replacing $m_0^2\to\CM^2$, $\alpha\to A$.  For a discussion of
contributions from $O(p^4)$ coefficients, see ref.
[\cite{GL}].

The $\Delta S=1,\ \Delta I=3/2\ \; K^+ \rightarrow \pi^+ \pi^0$ decay amplitude 
is 
proportional to the weak matrix element 
$$\langle  \pi^+ \pi^0 |{\left( {\bar s}d{\bar u}u+{\bar s}u{\bar u}d-
{\bar s}d{\bar d}d\right)}_{LL} |K^+ \rangle \; . \eqno(currentcurrent)$$
The four-fermion
operator is the $\Delta I=3/2$ component of the same 27-plet that
also contains the operator ${({\bar s}d{\bar s}d)}_{LL}$ (\eq{BK}). To 
$O(p^2)$ in ChPT, the operator is 
represented by
$$O_4=\alpha_{\scriptscriptstyle 27} \; r^{ij}_{kl}({\Sigma \partial^\mu 
\Sigma^\dagger})_i^{\ k}({\Sigma \partial_\mu \Sigma^\dagger})_j^{\ l}
\ , \eqno(rwOfour)$$
where the tensor $r^{ij}_{kl}$ has nonzero components 
$$r^{21}_{31}=r^{12}_{13}=r^{12}_{31}=r^{21}_{13}=
{1 \over 2}\ ,$$ 
$$r^{22}_{32}=r^{22}_{23}=-{1 \over 2}\eqno(ts)$$
(all other components vanish). The parameter 
$\alpha_{\scriptscriptstyle 27}$ is the same as in \eq{OOprime}. 
The aim is then to calculate the matrix element 
$\langle\pi^+\pi^0|O_4|K^+\rangle$ to one loop.

The lattice determination of $\langle\pi^+\pi^0|O_4|K^+\rangle$ was
compared with its real-world value in great detail at one loop in ChPT and
QChPT in ref. [\cite{GL}], and here we will only discuss what is new in
the partially quenched case.  

All attempts to compute this matrix element
on the lattice have been restricted to the mass-degenerate, quenched
theory, and moreover, all mesons are taken to be at rest [\cite{BS,It,JLQCD}].
The operator $O_4$ then inserts energy, implying that the 
values thus obtained are unphysical. 
All these systematic errors can be studied in ChPT. Deviations due to the
choice of unphysical masses and momenta already shows up at tree level
[\cite{BS}],
while all three systematic effects (including quenching) lead to
one-loop contributions different from those calculated for physical masses
and momenta in the unquenched theory [\cite{GL}]. 
In addition, at one loop one finds that there are power-like 
finite-volume corrections, which were also calculated in ref. [\cite{GL}].

The one-loop result for this unphysical matrix element
in a finite volume $L^3$ (with periodic boundary conditions) and with
$m_1=m_2=m_V$ (hence $M_\pi=M_K$, with $M_\pi$ the mass of the
$u\overline{d}$ meson)
for the partially quenched theory is, keeping only non-analytic corrections, 
$$\eqalignno{
\langle \pi^+ \pi^0 |O_4|K^+ \rangle^{p}_{unphys}=
{{24i
\alpha^{p}_{\scriptscriptstyle 27} M^2_\pi} \over {\sqrt{2}f^3_{p}}}
\Biggl(1
&-N {M_{1S}^2 \over (4\pi f_{p})^2}
\log{ M_{1S}^2 \over \Lambda^2_{p}}&(pq)\cr
&+{M^2_{11} \over (4 \pi f_{p})^2}\left[ -3\log{M^2_{11} \over
\Lambda^2_{p}}+F(M_{11} L)\right]
\Biggr)\ ,
}$$
up to corrections vanishing faster than any power of $L^{-1}$, where the
function $F$ is given by
$$ F(x)={17.827 \over {x}}+{12\pi^2 \over {x^3}}\;.\eqno(fv)$$
We note that the chiral logarithm due to the sea-quark loops does not 
diverge in the chiral limit $m_{V} \rightarrow 0$. Note also that
the result \eq{pq} does not depend on the parameters $m^2_0$ and
$\alpha$. For a discussion of contributions from $O(p^4)$ coefficients,
see ref. [\cite{GL}].

The above result can also be derived from the ``quark flow picture" 
[\cite{qflow,SharpeBK}], and the results of ref. [\cite{GL}].
First, take all quark masses equal, including the sea-quark mass.  
The quenched and unquenched results of ref. [\cite{GL}] then 
are the special cases obtained
from \eq{pq} by setting $N=0$ and $N=3$ respectively. The difference is
due to the sea-quark loops which are present in the unquenched case, but
not in the quenched case.  Therefore, if we now have $N$ instead of $3$
sea quarks, one obtains the correct result by multiplying the difference
between the unquenched and quenched results by $N/3$, yielding the first
chiral logarithm in \eq{pq}.  But now we also identified which of the
logarithms is due to sea-quark loops (the term proportional to $N$),
and we conclude that we obtain the partially quenched result for
$m_S\ne m_V$ by replacing $M_{11}^2\to M_{1S}^2$ in the term linear in $N$.
(Because of the structure of $O_4$ and the $O(p^2)$ vertices
from the effective lagrangian \eq{Lag}, there can
be at most one sea-quark on the loop.  Therefore, the result does not
depend on $M_{SS}$ and the parameters $m_0$ and $\alpha$, 
{\it cf.} \eq{Gij}.) 
Of course, one should keep in mind that the parameters $f_p$ and 
$\alpha_{\scriptscriptstyle 27}^p$ depend on $N$.

The power-like
finite-volume corrections in \eq{pq} are independent of $N$.  This 
follows from the fact that they originate from diagrams which do not
contain sea-quark loops [\cite{GL,BSW}].  

\subhead{\bf 5. Numerical examples}

In this section, we will give some numerical examples 
of the differences between the one-loop estimates for $B_K$ and
$\langle\pi^+\pi^0|O_4|K^+\rangle$ calculated in the partially quenched
theory and in the ``real world."  For $\langle\pi^+\pi^0|O_4|K^+\rangle$
we will also take the other systematic effects discussed in section 4
into account.  We will take values for the lattice parameters typical
of those used in recent numerical computations (for partially quenched
results for $B_K$, see refs. [\cite{Kilcup,KL}]; for 
$\langle\pi^+\pi^0|O_4|K^+\rangle$ we are not aware of any lattice data). 

The general strategy for our estimates is the same as in ref. [\cite{GL}]. 
We will set all $O(p^4)$ coefficients to zero. We choose the 
cutoffs $\Lambda$ (for the full theory) and $\Lambda_{p}$ (for the
partially quenched theory) to be 
1~GeV or  770~MeV, independent of one another. 
The sensitivity under a change in $\Lambda$ and 
$\Lambda_{p}$ is taken as an indication of the systematic 
error associated with our ignorance of the values of $O(p^4)$
coefficients.
For the real-world values of $f_\pi$, $m_\pi$ and $m_K$ we will use 
$f_\pi =132$~MeV, $m_\pi =136$~MeV and $m_K =496$~MeV. 

\bigskip 
\leftline{\it A. $B_K$}

$B_K$ in the full and partially quenched theories can be related 
by using \eq{QBK} above and Eq.~(36) of ref. [\cite{GL}]:
$${B^{phys}_K \over B^{p}_K}
={\alpha_{\scriptscriptstyle 27} \over \alpha^{p}_{\scriptscriptstyle 27} }
   \left( {f_{p} \over f}\right)^4 P\;, \eqno(Brel)
$$
where
$$P={1+H \over 1+{\tilde H}}\ , \eqno(P)$$
with $H=0.724$ (for $\Lambda =1$~GeV) and $H=0.417$ (for $\Lambda=770$~MeV)
is the numerical value of the relative one-loop correction for  
$B_K$ in the real world, and $\tilde H$ 
is the relative one-loop correction 
for the partially quenched theory in \eq{QBK}.  To one-loop accuracy,
$M_{12}^2$ can be replaced by $M_K^2$. 
The factor $P$ incorporates all one-loop corrections ($P=1$ at tree level). 

Since the ratios $f_p/f$ and $\alpha_{\scriptscriptstyle 27}/
\alpha_{\scriptscriptstyle 27}^p$ cannot be determined within ChPT, 
we will arbitrarily set them equal to one.
This constitutes one of the major uncertainties of our method. Furthermore,
we will take $\alpha=0$ for simplicity. 
In Table 1, we list the numerical values of $P$ for 
different combinations of cutoffs, choosing $f_{p}=f_\pi$, $\epsilon=1/2$,  
$M_{SS}=m_K$, $N=2$, for various choices of the parameters
$y=(m_0^2/3)/M_{SS}^2$ and $M^2_K$.

\bigskip
\bigskip
{\def\tablerule{\noalign{\hrule}}
\offinterlineskip
\centerline{
\vbox{\halign{%
\ \
\hfil#\hfil\tabskip=2em&\hfil#\hfil\tabskip=
3.3em&\hfil#\hfil\tabskip=2.8em&\hfil#\hfil\tabskip=2.6em&\hfil#\hfil
\strut\tabskip=2.6em&\hfil#\hfil\tabskip=.5em\cr
$M^2_K$&$y$&$P^{(1)}_{\
(1)}$&$P^{(0.77)}_{\ (0.77)}$&$P^{(1)}_{\ (0.77)}$&$P^{(0.77)}_{\
(1)}$\cr
\noalign{\vskip.1em \hrule height0pt depth1pt \vskip.5em}
 &0.5&1.02&0.98&1.19&0.84\cr
0.2&1&1.02&0.98&1.19&0.84\cr
 &$\infty$&1.01&0.96&1.17&0.83\cr
\noalign{\vskip.5em}\tablerule\noalign{\vskip.5em}
 &0.5&0.99&1.13&1.37&0.81\cr
0.4&1&0.98&1.11&1.35&0.81\cr
 &$\infty$&0.96&1.05&1.28&0.79\cr
\noalign{\vskip.5em}\tablerule\noalign{\vskip.5em}
 &0.5&1.11&1.73&2.10&0.92\cr
0.6&1&1.09&1.64&1.99&0.90\cr
 &$\infty$&1.03&1.42&1.73&0.84\cr
\noalign{\vskip.5em}\tablerule
}}
}
}
\bigskip
{\parindent=0pt
\it Table 1. The factor $P$, \eq{P},
for different values of $M^2_K$, $y$ and different combinations
of $\Lambda$ and $\Lambda_{p}$. Other parameters are fixed at
$f_{p}=f_\pi$, $\epsilon=1/2$,
$M_{SS}=m_K$ and $N=2$.
The superscript on $P$ denotes $\Lambda$ in {\rm GeV};
the subscript on $P$ denotes $\Lambda_{p}$ in {\rm GeV}.
$M^2_K$ is in ${\rm GeV}^2$. }

For $M^2_K =0.6$~GeV$^2$, values for $P$ as high as $2$ are obtained,
for $\Lambda_p=770$~MeV. However, for this cutoff, the meson mass is
probably too large for PQChPT to be reliable. Also, for $M_K^2=0.6$~GeV$^2$
and  $y=0.5$, $M_K^2/m_{\eta'}^2$ is of order one. We will therefore
concentrate on the two lower masses in the following discussion. 
For these values, we see that $P$ never differs from one by more 
than about $10\%$, if we choose $\Lambda=\Lambda_p$, which is 
equivalent to the assumption that the bare parameters of the
full and partially quenched theories are equal (this may not be
unreasonable for $N=2$).  However, if we
take the two cutoffs unequal, corrections can be as large as $20-30\%$.
The results  are fairly insensitive to changes in the value of $y$. 

\bigskip
\leftline{\it B. $K^+ \rightarrow \pi^+ \pi^0 $}

The $K^+ \rightarrow \pi^+ \pi^0 $ matrix elements in the full and partially 
quenched theories can be related by using \eq{pq} above and Eqs.~(43,87) of
ref. [\cite{GL}] (replacing $M_{11}^2$ by $M_\pi^2$ and $M_{1S}^2$ by
$M_{VS}^2$):
$$\langle \pi^+ \pi^0 |O_4 |K^+\rangle_{phys} =W\;
{\alpha_{\scriptscriptstyle 27} \over \alpha^{p}_{\scriptscriptstyle 27} }
\; \left( {f_{p} \over f} \right)^3 
\; {{m^2_K  
-m^2_\pi}\over 2M^2_\pi}\; \langle \pi^+ \pi^0 |O_4  
|K^+\rangle^{p}_{unphys}\ ,\eqno(fmr)$$
with
$$W={1+U \over 
1-N {M_{VS}^2 \over (4\pi f_{p})^2}
\log{ M_{VS}^2 \over \Lambda^2_{p}}
+{M^2_\pi \over (4 \pi f_{p})^2}\left[ -3\log{M^2_\pi \over
\Lambda^2_{p}}+F(M_\pi L)\right]} \ ,
\eqno(W)
$$
where $U=0.0888$ (for $\Lambda =1$~GeV) and $U=-0.0146$ (for $\Lambda=770$~MeV)
is the numerical value of the relative one-loop correction 
in the real world. At tree level, $W=1$. 

We list in Table~2a the numerical values of 
$W$ for different combinations of the cutoffs
$\Lambda$ and $\Lambda_p$, and for $f_{p} =f_\pi$, 
$M_{SS}=m_K$, $N=2$, 
several values of $M^2_\pi$ and 
a fixed volume $L^3$ such that $M_\pi L=6$ for $M_\pi^2=0.2$~GeV$^2$.
Here we may use $M_{VS}^2=(M_\pi^2+M_{SS}^2)/2$.
In Table~2b we list values of $W$ for infinite volume, with all other
parameters the same as in Table~2a.  

We see from Table 2 that one-loop corrections are always rather
large, even for relatively small meson masses, and that
sensitivity to the values of the cutoffs is significant. 
(For infinite volume, the corrections are not quite as big, 
{\it cf.} Table~2b.) This casts some doubt
on the accuracy of one-loop ChPT in estimating the factor $W$,
and one would expect that two-loop contributions are not small. 
As in the quenched case, the ``correction factor" $W$ is always
substantially smaller than one.  For a much more detailed discussion
of uncertainties inherent to our estimates of such ``correction factors,"
see ref. [\cite{GL}].

As can be seen from \eq{pq}, the partially quenched result is 
closer to the unquenched case (for which $N=3$, $M_{1S}=M_{11}$) than
to the quenched case.  The large deviations from the tree-level
value $W=1$ are mostly
due to the other systematic effects.  For instance, in the unquenched
theory, we find $W=0.56$ ($\Lambda=1$~GeV) and $W=0.57$ 
($\Lambda=770$~MeV) for $M_\pi^2=0.2$~GeV$^2$. 

\bigskip
\bigskip
{\def\tablerule{\noalign{\hrule}}
\offinterlineskip
\centerline{
\vbox{\halign{%
\ \
\hfil#\hfil\tabskip=2em&\hfil#\hfil\tabskip=
3.3em&\hfil#\hfil\tabskip=2.8em&\hfil#\hfil\tabskip=2.6em&\hfil#\hfil
\strut\tabskip=2.6em&\hfil#\hfil\tabskip=.5em\cr
$M^2_\pi$&$W^{(1)}_{\
(1)}$&$W^{(0.77)}_{\ (0.77)}$&$W^{(1)}_{\ (0.77)}$&$W^{(0.77)}_{\
(1)}$\cr
\noalign{\vskip.1em \hrule height0pt depth1pt \vskip.5em}
0.2&0.59&0.60&0.66&0.53\cr
0.4&0.54&0.60&0.66&0.49\cr
0.6&0.55&0.66&0.73&0.49\cr
\noalign{\vskip.5em}\tablerule
}}
}
}
\bigskip
{\parindent=0pt
\it Table 2a. The factor $W$
for different values of $M^2_\pi$ and different combinations
of $\Lambda$ and $\Lambda_{p}$. Other parameters are fixed at:
$f_{p}=f_\pi$,
$M_{SS} =m_K$, $N=2$ and $M_\pi L =6$ for $M_\pi^2=0.2$.
The superscript on $W$ denotes $\Lambda$ in {\rm GeV};
the subscript on $W$ denotes $\Lambda_{p}$ in {\rm GeV}.
$M^2_\pi$ is in ${\rm GeV}^2$. }
\bigskip
\bigskip
{\def\tablerule{\noalign{\hrule}}
\offinterlineskip
\centerline{
\vbox{\halign{%
\ \
\hfil#\hfil\tabskip=2em&\hfil#\hfil\tabskip=
3.3em&\hfil#\hfil\tabskip=2.8em&\hfil#\hfil\tabskip=2.6em&\hfil#\hfil
\strut\tabskip=2.6em&\hfil#\hfil\tabskip=.5em\cr
$M^2_\pi$&$W^{(1)}_{\
(1)}$&$W^{(0.77)}_{\ (0.77)}$&$W^{(1)}_{\ (0.77)}$&$W^{(0.77)}_{\
(1)}$\cr
\noalign{\vskip.1em \hrule height0pt depth1pt \vskip.5em}
0.2&0.68&0.71&0.78&0.62\cr
0.4&0.65&0.75&0.83&0.59\cr
0.6&0.68&0.90&0.99&0.62\cr
\noalign{\vskip.5em}\tablerule
}}
}
}
\bigskip
{\parindent=0pt
\it Table 2b. The factor $W$
for different values of $M^2_\pi$ and different combinations
of $\Lambda$ and $\Lambda_{p}$. Other parameters are fixed at:
$f_{p}=f_\pi$,
$M_{SS} =m_K$, $N=2$ and $M_\pi L =\infty$.
The superscript on $W$ denotes $\Lambda$ in {\rm GeV};
the subscript on $W$ denotes $\Lambda_{p}$ in {\rm GeV}.
$M^2_\pi$ is in ${\rm GeV}^2$. }
\bigskip

\subhead{\bf 6. Conclusion}

In this paper, we applied PQChPT to the calculation of the quark-mass
dependence of various quantities to one loop, restricting ourselves to
degenerate sea-quark masses.  For Goldstone boson
masses and decay constants, we extended results obtained earlier in
ref. [\cite{SharpePQ}]: we investigated the sensitivity 
of the chiral logarithms to the singlet
part of the $\eta'$ mass, $m_0$. Since
the ``remnants" of the $\eta'$ are an essential part of the fully
quenched theory, this is a natural question to ask in the partially
quenched theory.  We found that the coefficients of the chiral logarithms
for the Goldstone meson masses are sensitive to the value of $m_0$ 
for typical values of the sea-quark mass. 
The decay constants depend
on $m_0$ to a lesser extent, which is related to the fact that in
the limit of degenerate quark masses, the axial current does
not couple to the $\eta'$. 

We also calculated one-loop contributions to the chiral condensate,
$B_K$ and the $K^+\rightarrow\pi^+\pi^0$ decay amplitude, extending
earlier work in QChPT [\cite{SharpeBK,SharpeTASI,GL}] to the partially
quenched case. For $K^+\rightarrow\pi^+\pi^0$ we only considered the
case of degenerate valence-quark masses.  

We considered some numerical
examples of the comparison between $B_K$ and the $K^+\rightarrow\pi^+\pi^0$
decay amplitude in the real world and in partially quenched QCD with
values of the parameters typical of current lattice computations.
For $B_K$, we found that, for small enough meson masses, the ``correction
factor" may be very close to one, but with an uncertainty which could
be as large as $20-30\%$.  For $K^+\rightarrow\pi^+\pi^0$, we also
took into account that typical lattice computations are done at
unphysical (degenerate) valence-quark masses and external momenta, and 
we included the leading finite-volume corrections. We found
that the ``correction factor" is always much smaller than one (of
order one-half, with large uncertainties).  This is mostly due to
the unphysical choice of masses and momenta, and not to partial
quenching. 

\subhead{\bf Acknowledgements}

We thank Steve Sharpe for very useful comments on a preliminary
version of this paper.  
This work is supported in part by the US Department of
Energy (MG as an Outstanding Junior Investigator).

\references

\refis{MGMainz}
M.F.L.~Golterman, in {\it Chiral Dynamics: Theory and Experiment}, 
Proceedings of the
Workshop on Chiral Dynamics, Mainz, Germany, 1997, edited by A.M.~Bernstein,
D.~Drechsel and T.~Walcher, to be published (hep-ph/9710468).
 
\refis{BG}
C.W.~Bernard and M.F.L.~Golterman, Phys. Rev. D {\bf 49}, 486 (1994).

\refis{SharpePQ}
S.R.~Sharpe, Phys. Rev. D, in press (hep-lat/9707018).

\refis{Sharpe96}
S.R.~Sharpe, in {\it Lattice '96}, Proceedings of the International Syposium on 
Lattice Field Theory, St. Louis, USA, edited by C.~Bernard, M.~Golterman, M.~
Ogilvie and J.~Potvin 
[Nucl. Phys. B (Proc. Suppl.) {\bf 53}, 181 (1997)].

\refis{Morel}
A.~Morel,  J. Physique {\bf 48}, 111 (1987).

\refis{GL}
M.F.L.~Golterman and K.-C.~ Leung, Phys. Rev. D {\bf 56}, 5 (1997). 

\refis{SZ}
S.R.~Sharpe and Y.~Zhang,  Phys. Rev. D {\bf 53}, 5125 (1996).

\refis{BGQ}
C.W.~Bernard and M.F.L.~Golterman, Phys. Rev. D {\bf 46}, 853 (1992).

\refis{qflow}
S.R.~Sharpe, in {\it Lattice '89}, Proceedings of the International Syposium
on Lattice Field Theory, Capri, Italy, edited by N.~Cabibbo {\it et al.} 
[Nucl. Phys. B (Proc. Suppl.) {\bf 17} 146 (1990)].

\refis{KS}
G.~Kilcup {\it et al.}, Phys. Rev. Lett. {\bf 64}, 25 (1990).

\refis{Kilcup}
G.~Kilcup, Phys. Rev. Lett. {\bf 71}, 1677 (1993).

\refis{KL}
W.~Lee and M.~Klomfass, hep-lat/9608089.

\refis{SharpeBK}
S.R.~Sharpe, Phys. Rev. D {\bf 46}, 3146 (1992).

\refis{BGF}
C.W.~Bernard and M.F.L.~Golterman, in {\it Lattice '92}, Proceedings of the
International Syposium on
Lattice Field Theory, Amsterdam, The Netherlands, edited by J.~Smit and
P.~van~Baal 
[Nucl. Phys. B (Proc. Suppl.) {\bf 30}, 217 (1993)].

\refis{SharpeTASI}
S.R.~Sharpe, in {\it TASI 1994}, CP Violation
and the Limits of the Standard Model, Boulder, USA, edited by J.F.~Donoghue,
377 (1994).

\refis{DGH}
J.F.~Donoghue, E.~Golowich and B.R.~Holstein, Phys. Lett. B {\bf 119},
412 (1982).

\refis{BS}
C.W.~Bernard and A.~Soni, in {\it Lattice '88}, Proceedings of the
International Syposium on
Lattice Field Theory, Amsterdam, The Netherlands, edited by A.S.~Kronfeld 
and P.B.~Mackenzie
[Nucl. Phys. B (Proc. Suppl.) {\bf 9}, 155 (1989)]. 

\refis{It}
M.B.~Gavela {\it et al.}, Nucl. Phys. B {\bf 306}, 677 (1988).

\refis{JLQCD}
N.~Ishizuka, for JLQCD, in {\it Lattice '97}, Proceedings of the
International Syposium on
Lattice Field Theory, Edinburgh, Scotland, edited by C.~Davies 
{\it et al.}, to be published.

\refis{BSW}
J.~Bijnens, H.~Sonoda and M.B.~Wise, Phys. Rev. Lett. {\bf 53},
2367 (1984).

\refis{KV}
L.~Venkataraman and G.~Kilcup, hep-lat/9711006.

\refis{BGscat}
C.W.~Bernard and M.F.L.~Golterman, Phys. Rev. D {\bf 53} 476 (1996).

\refis{CP}
G.~Colangelo and E.~Pallante, hep-lat/9708005. 

\endreferences

\vfil
\bye